\documentclass[aps,notitlepage,a4paper,superscriptaddress,11pt]{article}
\usepackage{cmbright}
\usepackage[OT1]{fontenc}
\usepackage{titlesec,titletoc,authblk,abstract,latexsym,booktabs,amsmath,cleveref,fancyhdr,wrapfig,array,amsfonts,amssymb,geometry,appendix,cite}
\usepackage{xcolor}

\newcommand{\be}{\begin{equation}}
\newcommand{\ee}{\end{equation}}
\newcommand{\bea}{\begin{eqnarray}}
\newcommand{\eea}{\end{eqnarray}}
\newcommand\blfootnote[1]{%
  \begingroup
  \renewcommand\thefootnote{}\footnote{#1}%
  \addtocounter{footnote}{-1}%
  \endgroup
}

\usepackage{graphicx,lipsum}
\usepackage{caption}
\usepackage{subfigure}
\usepackage{amsmath}
\usepackage{amsfonts}
\usepackage{amssymb}
\usepackage{mathtools}
\numberwithin{equation}{section}
\usepackage{titlesec}

\numberwithin{subcase}{case}
\usepackage{framed}
\allowdisplaybreaks
\usepackage[nottoc]{tocbibind}
\usepackage[free-standing-units]{siunitx}
\usepackage{helvet}
\usepackage{url}
\usepackage{titletoc}
\usepackage{blindtext}
\usepackage[gen]{eurosym}

\title{Solving Non-hermitian Dirac equation in the presence of PDM and local Fermi velocity }
\author{Rahul Ghosh}
\affil{Physics Department, Shiv Nadar University, Gautam Buddha Nagar, \\ Uttar Pradesh 201314, India}

\begin{document}

\maketitle

\begin{abstract}
Abstract: We present a new approach to study a class of non-Hermitian (1+1)-dimensional Dirac Hamiltonian in the presence of local Fermi velocity. We apply the  well known Nikiforov-Uvarov method to solve such a system. We discuss applications and explore the solvability of both $\mathcal{PT}$-symmetric and non-$\mathcal{PT}$ symmetric classes of potentials. In the former case we obtain the solution of a harmonic oscillator in the presence of a linear vector potential while in the latter case we solve the shifted harmonic oscillator problem.
\end{abstract}
\blfootnote{E-mails: rg928@snu.edu.in}

\section{Introduction}
Dirac equation has been hailed as one of the finest achievements in the history of physics. It seeks to combine the basic tenets of special relativity with the principles of quantum mechanics \cite{dirac, tha1992}. A glance at the literature will reveal that its applications are of numerous considerations ranging from addressing problems in condensed matter physics \cite{kat} and high energy physics \cite{khr} to many other branches of physics, including the recently emerged field of graphene \cite{nov, cast, gall,jakub22,fer2020}. 

From the theoretical perspective, there has been some interest in the applications of Dirac equation in the non-hermitian domain of quantum mechanics. Among major issues that have come up is the optical realization of non-hermitian Dirac Hamiltonian in the space-dependent effective mass background \cite{lon2010}. Another one is concerned with graphene when dissipative forces are operating in the presence of an external fixed force \cite{flo2021}. 

In recent times, the subclass of nonhermitian quantum systems incorporating the parity ($\mathcal{T}$)-time ($\mathcal{T}$ symmetry has proved to be an active area of research 
\cite{ben1998, bag2000, moi, baga2016, elg, cruz,ahm2001}. $\mathcal{PT}$-symmetric Hamiltonians support real or complex
conjugate pairs of energy eigenvalues under certain conditions related to enforcing $\mathcal{PT}$ to be exact or a spontaneously broken symmetry. In this context, we point out that heavy Dirac particles could be tackled in the presence of a non-$\mathcal{PT}$-symmetric potential \cite{ard2009}. Alhaidari has also studied the case of non-$\mathcal{PT}$ potentials accounting for localized and/or continuum states \cite{alh2013}. Other works on non-$\mathcal{PT}$ potentials include \cite{bag2020} and references cited therein.

For systems exhibiting $\mathcal{PT}$-symmetry, let us note that the operations of parity $ \mathcal{P}$ and time reversal $\mathcal{T}$ transform as $ \mathcal{P} : x \rightarrow -x$ and $\mathcal{T} : i \rightarrow -i$. Consequently their joint operation has the role
\begin{gather}
\mathcal{PT} : x \rightarrow -x \quad \text{and} \quad \mathcal{PT}: p \rightarrow p
\end{gather}
For the Hamiltonian to be $\mathcal{PT}$-symmetric, it should commute with the $\mathcal{PT}$ operator namely, 
$ [\hat{H}, \hat{\mathcal{P}}\hat{\mathcal{T}}]=0 $. 

The idea of spatially varying Fermi velocity has also attracted attentions after a gap formation was noticed in graphene physics \cite{gui2008}. In a recent  study, it was shown that, for definite spatial dependencies of Fermi velocity, Dirac particles experience an effective magnetic field in a nonuniform honeycomb lattice \cite{dow2016, ley2020}. It bears mention that the implementation of spatially varying Fermi velocity in the Dirac equation was initially suggested by Downing and Portnoi in \cite{dow2017} to enquire how coordinate fluctuations of the Fermi velocity can lead to localization effects in graphene.

On the other hand, when the system is non-hermitian, the position-dependent effective mass Dirac equation has been addressed by utilising the Lorentzian 2-vector potential \cite{jia2006, dut2009} . In a similar vein, the Klein-Gordon equation has also been examined for both non-PT/non-hermitian instances in \cite{arda10}. Studies on the Schrödinger equation with effective mass include the following: deformed shape invariance and exactly solvability \cite{bagchi05}, generating associated potentials from the kinetic energy \cite{bagchi04}, quasi-particles with indefinite effective mass that depend on both position and excitation \cite{znojil12}, and exact solution for the pseudoharmonic potential obtained for an arbitrary angular momentum \cite{sever08}. For more of the theory of position-dependent mass systems in both, classical and quantum mechanics see \cite{bagchi2008, oscar20}. 

Mixing of $\mathcal{PT}$-symmetric vector, scalar and pseudoscalar potentials for time-independent Dirac equation has pointed to the existence of real energies in the system \cite{dut2009}. In this paper, a (1+1)-dimensional position-dependent mass (PDM) Dirac equation is investigated by considering the effects of a local Fermi velocity (LFV) to generate a large class of associated solvable potentials \cite{rah2022}. The idea of LFV was also explicitly used in \cite{mus2013}.
Our purpose here is to exploit the Nikiforov–Uvarov (NU) technique \cite{nik1988} to explore and solve Dirac equation in a non-hermitian setting by implementing a local Fermi velocity in the underlying Hamiltonian 
Subsequently, the study of both types of cases are taken up, one related to the Hamiltonian being $\mathcal{PT}$-symmetric while the other addressing a non-$\mathcal{PT}$ system. It may be noted that in \cite{rah2022,mus2013} the hermitian cases have been already addressed.

The organization of this paper is as follows: we start with a general Dirac Hamiltonian in the background of space-dependent mass and Fermi velocity; in Section 2, the necessary conditions to realize the $\mathcal{PT}$-symmetry are discussed; in Section 3, the structure of the Dirac equation is decoupled to facilitate application of the NU method to arrive at a generalized form of a second-order differential equation; in Section 4, we introduce a transformation to the complex plane to solve the latter equation; in Section 5, applications to $\mathcal{PT}$ and non-$\mathcal{PT}$ models are considered; finally, Section 6 concludes with a summary.

\section{Dirac Equation with spatial variation of mass and Fermi velocity}
We first focus on the hermitian form of the Dirac Hamiltonian $H_D$. To this end, we start with the $(1+1)$-dimensional representation \cite{jun2020,dut2009}
\begin{equation} \label{mainDirac}
H = v_f p_x \sigma_3 + (S(x)+m v_f^2) \sigma_2 + W(x) \sigma_1 + V(x) \mathcal{I}
\end{equation}
where $m$ is the mass of the spin-$\frac{1}{2}$ particle, $v_f$ is the Fermi velocity and $\mathcal{I}$ is the block-diagonal unit matrix.  The standard expressions of the Pauli matrices are known to be
\begin{gather}
\sigma_1 = \left( \begin{array}{cc} 0 & 1  \\ 1 & 0  \end{array} \right), \quad \sigma_2 = \left( \begin{array}{cc} 0 & -i  \\ i & 0  \end{array} \right), \quad \sigma_3 = \left( \begin{array}{cc} 1 & 0  \\ 0 & -1  \end{array} \right) 
\end{gather}
The associated governing potentials appearing in (2.1) are the vector potential $V(x)$, scalar potential $S(x)$ together with the pseuodoscalar contribution $W(x)$. From the supersymmetric quantum mechanics point of view, $W(x)$ acts as the superpotential of the system \cite{bag2021}.


We let the Fermi velocity to be space-dependent \cite{per2009,rah2022} for the (1+1)-dimensional hererostructure. In other words, we treat the Fermi velocity to be a local variable (LFV). In the presence of LFV and PDM, we are thus led to the following extended form of the Dirac Hamiltonian
\begin{equation} \label{HDvf}
H_D = \sqrt{v_f(x)} p_x \sqrt{v_f(x)} \sigma_3 + \Big( S(x)+ m v^2(x) \Big) \sigma_2 + W(x) \sigma_1 + V(x) \mathcal{I} 
\end{equation}
where $m \equiv m(x)$ is a real function of x. Observe that the introduction of LFV does not affect the hermiticity of the Hamiltonian. 
With (\ref{HDvf}) for $H_D$ at hand, we re-express it in the form 

\begin{equation}
H_D = \sqrt{v_f(x)} p_x \sqrt{v_f(x)} \sigma_3 + R(x) \sigma_2 + W(x) \sigma_1 + V(x) \mathcal{I} 
\end{equation}
where we have defined $R(x)$ through the relation

\begin{equation}\label{H_D0}
    S(x)+ m v_f^2(x)= R(x)
\end{equation}
The spatial dependence in the mass is implicit in $R(x)$.
\section{Decoupling the Dirac Equation}
Let us observe that the time-independent Dirac equation satisfies
\begin{gather} \label{TIDiric}
    H_D \Psi(x)= E \Psi(x)
\end{gather}
where $\Psi(x)$  represents a two-component spinor wave function namely, $(\psi_1(x) \quad \psi_2(x))^T$. 
The matrix equation (3.1) in explicit form reads
($\hbar=1)$,
\begin{gather}
      \left( \begin{array}{cc} -i \sqrt{v_f(x)} \partial \sqrt{v_f(x)}+V(x)  &  W(x) -iR(x)  \\  W(x)+ iR(x)   & i \sqrt{v_f(x)} \partial \sqrt{v_f(x)}+V(x)   \end{array} \right) \left( \begin{array}{cc} \psi_1(x)    \\  \psi_2(x)   \end{array} \right) = E \left( \begin{array}{cc} \psi_1(x)    \\  \psi_2(x)  \end{array} \right)
\end{gather}
and implies a pair of coupled equations 
\begin{gather} \label{coupledpsi}
 -i \sqrt{v_f} \partial \sqrt{v_f} \psi_1 +(W-iR) \psi_2 = (E-V)\psi_1   \\
 i \sqrt{v_f} \partial \sqrt{v_f} \psi_2 +(W+iR) \psi_1 = (E-V)\psi_2 
\end{gather}

To tackle these equations, we make use of the so-called Nikiforov-Uvarov (NU) method \cite{nik1988}. The procedure is described in detail in the Appendix A. We choose the following set of transformations to map the spinor wavefunction $\Psi(x)$ to a new set of representation $\Sigma(x)$, $ \Sigma(x)=(\xi_1(x) \quad \xi_2(x))^T $, where the elements are
\begin{gather} \label{sigmas}
    \xi_1(x)=\frac{\sqrt{v_f(x)}\psi_1(x)}{W(x)-iR(x)}   \quad \text{and} \quad \xi_2(x)=\frac{\sqrt{v_f(x)}\psi_2(x)}{W(x)+iR(x)}
\end{gather}
$\xi_1$ and $\xi_2$ also satisfy another pair of coupled equations but can be uncoupled easily following the approach of \cite{rah2022}. For our need, we write down the second-order differential equation for  $\xi_1(z)$ 
  
 \begin{gather}   \label{SEupperx}
    \xi_1''+ \frac{v'_f-i f}{ v_f}\xi_1'+ \frac{iv_f(V'-f')+E^2+E(f-2V)-(W^2+R^2)+V^2-Vf }{ v_f^2}\xi_1  =0
\end{gather}
where $f$, function of $x$, is given by
\begin{gather}
    f(x)=i v_f(x) \frac{W'(x)-iR'(x)}{W(x)-iR(x)} 
\end{gather}
and the primes now denote derivatives with respect to $x$. Note that a similar differential equation also exists for $\xi_2$.

To facilitate comparison with the NU-equation as given in (\ref{NUequation}) we rewrite (\ref{SEupperx}) in the following form 
\begin{gather} \label{SEupperz}
 \ddot{\xi_1} + \frac{(\dot{v}_f + f)}{v_f}\dot{\xi_1}+ \frac{W^2+R^2-V^2+ V f-v_f(\dot{V}-\dot{f})-E^2-E(f-2V)}{v_f^2}\xi_1 =0
\end{gather}
where the dot stands for the derivative with respect to $z$ and $z$ is defined by $z = -ix$, as was done in  \cite{has2013,kempf95}. The function $f$ is transformed to
\begin{gather} \label{f(z)}
f(z)=v_f(z) \frac{\dot{W}(z)-i\dot{R}(z)}{W(z)-iR(z)}   
\end{gather}
 Equation (\ref{SEupperz}) is the central equation to be explored for application to specific models.

\section{Applications}

\subsection{$\mathcal{PT}$-symmetric Case: Linear LVF and Linear Vector potential }

As a simple illustration, we choose the function $R(z)$ to be zero 
and take pseudoscalar and vector potentials to be
\begin{gather}
 W(z)= \omega \qquad \text{and }\quad V(z)= a+ b z
\end{gather}
 where the constants $\omega$, $a$ and $b$ are real. We assume for the LFV to be the linear form 
\begin{gather}
v_f(z)= 1+ \gamma z
\end{gather} 
To be $\mathcal{PT}$-symmetric, the coordinate-dependent Fermi velocity, and the different potentials in (\ref{H_D0}) need to fulfill the conditions 
\begin{gather} \label{PTcond}
v^*_f(x)=v_f(-x), \quad S^*(x) = S(-x), \quad V^*(x) = V(-x), \quad  W^*(x) = -W (-x)
\end{gather}
The possibility of the scalar potential $S(x)$ to be complex has been considered in the literature \cite{dut2009, yes2014}. In our case we have from (2.5) the following constraint 

\begin{gather}
  S_R (x) =  \frac{\gamma^2 x^2 - 1}{2\gamma x} S_I (x)
 \end{gather}
where $S_R$ and $S_I$ are the real and imaginary components of S(x). 
The above restriction holds for any arbitrary mass function $m(x)$ which may be chosen appropriately.

 

 These choices make the f-function to be zero i.e $f(z)=0$ and $\xi_1$ satisfying the simple differential equation 
\begin{gather} \label{upperNU1}
     \ddot{\xi_1} + \frac{ \gamma}{(1+\gamma z)}\dot{ \xi_1}+ \frac{-b^2 z^2- c z -d}{(1+\gamma z)^2}\xi_1 =0
\end{gather}
where the quantities $c$ and $d$ are

\begin{equation}
    c = 2ab+b\gamma-2bE, \qquad d= E^2+a^2+b-2aE-\omega^2 
\end{equation}

Comparing $(4.5)$ with the NU-equation $(\ref{NUequation})$ we have the correspondences
\begin{gather}
    \Tilde{\tau}(z)=  \gamma, \qquad \sigma(z)= 1+\gamma z, \qquad    \qquad \Tilde{\sigma}(z)=-b^2 z^2 - c z - d
\end{gather}
while the function $\pi (z)$ appearing in equation $(\ref{NUpi})$ is reduced to 
\begin{gather}  \label{pi2}
    \pi(z)=  \pm \sqrt{\left( b z+\frac{c+k\gamma}{2b} \right)^2+\left[ k+d-\frac{(c+k\gamma)^2}{4b^2}  \right] }
\end{gather}
We can fix the value of $k$ such that the second term inside the square root appearing in the square-parenthesis of (4.8) becomes zero leading to the following values of $k_\pm$
\begin{gather} \label{k2}
    k_\pm= \frac{2 b^2- b \gamma (2 a-2 E+\gamma )\pm 2b \sqrt{(b-a\gamma + E\gamma )^2-\gamma ^2 \omega ^2}}{\gamma ^2}
\end{gather}

The requirement of negative derivative of $\tau(z)$ implies that we need to consider the negative sign in the above and hence focus on $k_+$. Thus we arrive at the forms
\begin{gather}
    \pi(z)=- bz-\frac{c+k_+ \gamma}{2b}, \quad \tau(z)=-2 b z-\frac{c+k_+ \gamma}{b}+\gamma
\end{gather}

The eigenvalues can be obtained from $(\ref{lamda})$ and are given by 
\begin{gather}
E=( n+1 )\frac{\gamma}{2}  + \frac{\omega ^2}{2(n+1) \gamma }  \quad \text{\mbox{where} } n=0,1,2...
\end{gather}
The above result is new and of interest. It represents a combination of the harmonic oscillator's energies and a contribution coming from the linear vector potential. 

For completeness, let us furnish the results for the  $\theta(z)$ and the weight function $\rho(z)$ defined in (\ref{theta}) and (\ref{rho}). These are given by
\begin{gather}
    \theta(z)=e^{-a z} (1+\gamma z)^{\frac{\delta_0}{\gamma}}, \qquad \rho(z)=e^{-2a z} (1+\gamma z)^{\frac{2\delta_0}{\gamma}}
\end{gather}
where $\delta_0= a-\frac{c+k_+ \gamma}{2b}$. Therefore with the help of Rodrigues formula $(\ref{rodformula})$ the function $y_1n(z)$  turns out to be 
\begin{gather} 
    y_{1n}(z)= B_n e^{2a z} (1+\gamma z)^{\frac{-2\delta_0}{\gamma}}   \frac{d^n}{dz^n} \left[  e^{-2a z} (1+\gamma z)^{\frac{2\delta_0+n}{\gamma}}  \right]
\end{gather}


Therefore the total solution of equation $(\ref{upperNU1})$, derived by  $\xi_{1n}(z)=\theta(z)y_{1n}(z)$ from (\ref{Sigma}), reads as
\begin{gather}
    \xi_{1n}(z)= B_n e^{a z} (1+\gamma z)^{\frac{-\delta_0}{\gamma}}   \frac{d^n}{dz^n} \left[ e^{-2a z} (1+\gamma z)^{\frac{2\delta_0+n}{\gamma}}  \right]
\end{gather}
this solution can be rewritten in terms of associated Laguerre Polynomials $ L_\eta^\mu(s) $
\begin{gather}
\xi_{1n}(z)=B_n \gamma^n n! e^{-a z} (1+\gamma z)^{\frac{\delta_0}{\gamma}} L_n^{\frac{2\delta_0}{\gamma}}\left ( \frac{2 a }{\gamma} (1+\gamma z) \right)
\end{gather}
where $ L_\eta^\mu(s)=\frac{s^{-\mu} e^{s}}{n!} \frac{d^n}{dx^n} \left[  s^{\eta+\mu}e^{-s}  \right] $. Returning to $x$-space the above turns out to be 
\begin{gather}
\xi_{1n}(x)=B_n \gamma^n n! e^{ia x} (1-i\gamma x)^{\frac{\delta_0}{\gamma}} L_n^{\frac{2\delta_0}{\gamma}}\left ( \frac{2 a }{\gamma} (1-i\gamma x) \right)
\end{gather}

Finally we can write the complete upper component of the spinor wave function $\psi_{1n} (x)$, by using (\ref{sigmas}), as follows
\begin{gather}
    \psi_{1n}(x)= B_n \omega \gamma^n n! e^{ia x} (1-i\gamma x)^{\frac{2\delta_0-\gamma}{2\gamma}} L_n^{\frac{2\delta_0}{\gamma}}\left ( \frac{2 a }{\gamma} (1-i\gamma x) \right)
\end{gather}
and the lower component derived from (\ref{coupledpsi})
\begin{align}
 \psi_{2n}(x)=  \frac{B_n \gamma^n n! e^{ia x}  (1-i \gamma x )^{\frac{2\delta_0+\gamma}{2\gamma}} n!}{2} 
 & \Bigg[ \Big( \frac{ (1+n)^2\gamma^2+\omega^2}{(1-i\gamma x)(n+1)\gamma}-4a \Big)
 L_n^{\frac{2\delta_0}{\gamma }} \left ( \frac{2 a }{\gamma} (1-i\gamma x) \right) \notag \\ 
&  -4a  L_{n-1}^{\frac{2\delta_0+\gamma }{\gamma }} \left ( \frac{2 a }{\gamma} (1-i\gamma x) \right) \Bigg]   
\end{align}
with the above forms for $\psi_{1n}(x)$ and $\psi_{2n}(x)$ it can be shown that the Dirac spinor $\Psi(x)$ is normalizable.

\subsection{Non-$\mathcal{PT}$ symmetric Case: Linear LFV and Linear Scalar potential}
Let us employ the relationship $V(z)=\frac{f(z)}{2}$ as proposed in \cite{jia2006}, and choose again a linear form of the LFV 
\begin{gather}
v_f(z)= 1+\gamma z
\end{gather}
where $\gamma$ is a real parameter. Assuming pseudosscalar and scalar potentials to be given by
\begin{gather} \label{R(z)}
 W(z)=\alpha, \qquad \quad R(z)=i\beta z
\end{gather}
where the couplings $\alpha$ and $\beta$ are real , we note that in the $x$-space, $v_f(x)$ and $R(x)$ read
\begin{gather} \label{R(x)}
  v_f(x)= 1-i \gamma x, \qquad  \qquad R(x)=\beta x  
\end{gather}
 implying that we have at hand a non-$\mathcal{PT}$ problem since $R(x)$ is entirely real although $v_F (x)$ is $\mathcal{PT}$-symmetric. 
 
 Using (\ref{R(z)}) and (\ref{R(x)}), we can exploit the relation (2.5) to write down a set of consistent equations involving the mass function $m(x)$
 
 \begin{equation}
 S_R (x) = \beta x - (1-\gamma^2x^2) m(x), \quad S_I (x) = 2\gamma x m(x)
 \end{equation}
 where $S_R$ and $S_I$ are the real and imaginary components of the scalar potential $S(x)$. These in turn provide the relationship 
 
 \begin{equation}
     S_R (x) = \beta x + \frac{\gamma^2 x^2 - 1}{2\gamma x} S_I (x)
 \end{equation}

 Using specific forms of $S_R$ and $S_I$ we can generate a variety of forms of $m(x)$. For instance, if we take $m (x) = 0$ as in the case of graphene, then the scalar function turns out to be an entirely real function.

 The form (\ref{f(z)}) implies that the f-function, as well as the vector potential, is constant
\begin{gather}
f(z)=\gamma = 2 V(z)
\end{gather}
Equation $(\ref{SEupperz})$ then takes the form
\begin{gather} \label{upperNU2}
     \ddot{\xi_1} + \frac{2 \gamma}{(1+\gamma z)}\dot{ \xi_1}+ \frac{\epsilon^2-\beta^2 z^2}{(1+\gamma z)^2}\xi_1 =0
\end{gather}
where $\epsilon^2= \alpha^2+\frac{\gamma^2}{4}-E^2$ 


Comparing with the NU-equation $(\ref{NUequation})$  gives 
\begin{gather}
    \Tilde{\tau}(z)=2\gamma, \qquad \sigma(z)=1+\gamma z, \qquad    \qquad \Tilde{\sigma}(z)=-\beta^2 z^2 +\epsilon^2
\end{gather}
As a result, the corresponding representations $\pi(z)$ and $\tau(z)$ are given by  
\begin{gather}  \label{pitau}
    \pi(z)= -\frac{\gamma}{2} \pm \sqrt{\frac{\gamma^2}{4}-(\epsilon^2-\beta^2 z^2)+k(1+\gamma z))}, \quad \tau(z)=\Tilde{\tau}(z)+2\pi(z)
\end{gather}
From the condition discussed in the appendix the possible values of $k$, for which the expression under the square root sign becomes square of a polynomial, appear as
\begin{gather}
    k_\pm=2\alpha^2 \pm 2\alpha E
\end{gather}

To get a negative derivative of $\tau(x)$  we need to restrict  only to the negative sign of the square root appear in $(\ref{pitau})$. Further, keeping to $k_+$ value is relevant. This leads to the forms
\begin{gather}
    \pi(z)=-\beta z-\frac{k_+}{2\alpha}-\frac{\gamma}{2},
\qquad \quad 
  \tau(z)=-2\beta z-\frac{k_+}{\alpha}+\gamma
\end{gather}

The energy eigenvalues  can be obtained from $(\ref{lamda})$ and are
\begin{gather}
E_n=(n+\frac{1}{2})\gamma-\alpha, \qquad n=0,1,2...
\end{gather}
 We can easily identify the above energy levels with that of a harmonic oscillator with a constant shift. Thus we find real egenvalues for the non-$\mathcal{PT}$ potential $R(x)$ why this example is one of our important findings where we started with an non-$\mathcal{PT}$ symmetric potentials but still we have real eigenvalues. 

The functions $\theta(z)$ and $\rho(z)$ read, from (\ref{theta}) and (\ref{rho}),
\begin{gather}
    \theta(z)=e^{-\alpha z} (1+\gamma z)^{\frac{\delta_1}{\gamma}}, \quad \rho(z)=e^{-2\alpha z} (1+\gamma z)^{\frac{\delta_2}{\gamma}}
\end{gather}
where $\delta_1= \alpha-\frac{k_+}{2\alpha}-\frac{\gamma}{2}$ and $\delta_2= 2\alpha-\frac{k_+}{\alpha}$. Concerning the function $y_{1n}(z)$ defined in the Appendix A, Rodrigues formula $(\ref{rodformula})$ provides us with
\begin{gather} 
    y_{1n}(z)= A_n (1+\gamma z)^{-\frac{\delta_2}{\gamma}}e^{2\alpha z}   \frac{d^n}{dz^n} \left[  (1+\gamma z)^{n+\frac{\delta_2}{\gamma}}e^{-2\alpha z}  \right]
\end{gather}
Hence, $\xi_{1n}(z)$, as given by $\theta(z)y_{1n}(z)$ (see A.2), assumes the form
\begin{gather}
    \xi_{1n}(z)= A_n (1+\gamma z)^{\frac{\delta_1-\delta_2}{\gamma}}e^{\alpha z}   \frac{d^n}{dz^n} \left[  (1+\gamma z)^{n+\frac{\delta_2}{\gamma}}e^{-2\alpha z}  \right]
\end{gather}
The above can be rewritten in terms of the associated Laguerre Polynomials $ L_\eta^\mu(s) $
\begin{gather}
\xi_{1n}(z)=A_n \gamma^n n! (1+\gamma z)^{\frac{\delta_1}{\gamma}}e^{-\alpha z} L_n^{\frac{\delta_2}{\gamma}}\left ( \frac{2\alpha}{\gamma} (1+\gamma z) \right)
\end{gather}
where $ L_\eta^\mu(s)=\frac{s^{-\mu} e^{s}}{\eta!} \frac{d^\eta}{ds^\eta} \left[  s^{\eta+\mu}e^{-s}  \right] $. 

In terms of the x-variable, the above form transforms to
\begin{gather}
    \xi_{1n}(x)=A_n \gamma^n n! (1-i\gamma x)^{\frac{\delta_1}{\gamma}}e^{i\alpha x} L_n^{\frac{\delta_2}{\gamma}}\left ( \frac{2\alpha}{\gamma} (1-i\gamma x) \right)
\end{gather}
and we have the full expression for the upper component $ \psi_{1n}(x)$
\begin{gather}
    \psi_{1n}(x)=A_n \alpha \gamma^n n! (1-i\gamma x)^{\frac{2\delta_1+\gamma}{2\gamma}}e^{i\alpha x} L_n^{\frac{\delta_2}{\gamma}}\left ( \frac{2\alpha}{\gamma} (1-i\gamma x) \right)
\end{gather}
For completeness we give the expression of the lower component $ \psi_{2n}(x)$ 
\begin{align}
 \psi_{2n}(x)=  A_n \gamma^n n! e^{i\alpha x}  (1-i \gamma x )^{\frac{\delta_1-\gamma}{2\gamma}} n!  
 & \Bigg[ \Big( \gamma +n \gamma-\alpha  (2-i\gamma x)+\delta_1 \Big)
 L_n^{\frac{\delta_2}{\gamma }} \left ( \frac{2\alpha}{\gamma} (1-i\gamma x) \right) \notag \\ 
& -2 \alpha  (1-i \gamma x) 
L_{n-1}^{\frac{\delta_2+\gamma }{\gamma }} \left ( \frac{2\alpha}{\gamma} (1-i\gamma x) \right) \Bigg]   
\end{align}
as determined from (\ref{coupledpsi}).

\section{Summary}
To sum up, we explored a class of 1-dimensional non-hermitian Dirac Hamiltonian against the background of spatially-dependent mass and Fermi velocity. We took the Dirac equation in the standard form that exists in the literature and carried out its decoupling so that a particular type of the coupled pair could be compared with NU-form. The resultant consistency equation involved the superpotential which is the basic equation we sought to explore. In search of a viable solution we made use of the NU-method by making a suitable transformation of the spinor wave function. As applications we applied our procedure to two types of potentials, one being $\mathcal{PT}$ and the other of a non-$\mathcal{PT}$ type. For the LFV we adopted a linear form. In the first case a combination of the harmonic oscillator and a linear vector potential emerges as a viable solution while in the second example, the solution of a shifted oscillator is the outcome. In the second case because its spectrum is all-real, we could conclude that $\mathcal{PT}$-symmetry is not a necessary condition for a nonhermitan system to exhibit real spectra. In both the cases we derived closed-form solutions of the wave functions as given by the associated Lagurre polynomials. Finally, the techniques of NU-method has been outlined in the Appendix A.

\section{Acknowledgment}
I sincerely thank my supervisor Prof. Bijan Bagchi for the valuable guidance. I also thank Shiv Nadar University for the grant of senior research fellowship.

\appendix
\section{Nikiforov-Uvarov Technique}
Here we will provide some technical details of the technique \cite{moh2012} we used. We have a Schr\"{o}dinger type equation in some coordinate which is given by
\begin{gather} \label{NUequation}
    \frac{d^2 \Sigma}{dz^2}+\frac{\Tilde{\tau}(z)}{\sigma(z)}\frac{d\Sigma}{dz}+\frac{\Tilde{\sigma}(z)}{\sigma^2(z)}\Sigma=0
\end{gather}
where  $\Tilde{\tau}(z)$ is first degree polynomial of $z$ while  $\sigma(z)$ and $\Tilde{\sigma}(z)$ are at most second degree polynomial. Most interesting thing is that $z$, $\Tilde{\tau}(z)$, $\sigma(z)$ and $\Tilde{\sigma}(z)$ can be real as well as complex variable. By splitting $\Sigma(z)$ into two independent functions 
\begin{gather} \label{Sigma}
  \Sigma(z)=\theta(z)y(z)  
\end{gather}
After replacing the above we have  $(\ref{NUequation})$ as in the form
\begin{gather} \label{NUequation2}
   \frac{d^2 y}{dz^2} + \Big(2\frac{\theta'}{\theta}+\frac{\Tilde{\tau}}{\sigma} \Big) \frac{dy}{dz} + \Big(\frac{\theta''}{\theta}+\frac{\theta'}{\theta}\frac{\Tilde{\tau}}{\sigma}+\frac{\Tilde{\sigma}}{\sigma^2}  \Big)y=0
\end{gather}
Now we introduce $\pi(z)$ as follows  
\begin{gather} \label{theta}
 \frac{\theta'(z)}{\theta(z)}=\frac{\pi(z)}{\sigma(z)}   
\end{gather}
 and $\tau(z)$ as 
\begin{gather}\label{taupi}
    \tau(z)=\Tilde{\tau}(z)+2\pi(z)
\end{gather}
where $\pi(z)$ and $\tau(z)$ are polynomials of degree at most one.

Further, assuming the coefficient of $y$ of $(\ref{NUequation2})$ as $\frac{\Bar{\sigma}(z)}{\sigma^2(z)}$ which can be expressed as 
\begin{gather} \label{sigmabarpi}
   \Bar{\sigma}(z)= \Tilde{\sigma}(z)+ \pi^2(z)+ \pi(z)(\tau(z) - \sigma'(z)) + \pi'(z)\sigma(z)
\end{gather}

As a consequence of the algebraic transformations mentioned above, the form of $(\ref{NUequation})$ is protected in a systematic way and look like 
\begin{gather}
    \frac{d^2 y}{dz^2} + \frac{\tau}{\sigma}  \frac{dy}{dz} + \frac{\Bar{\sigma}}{\sigma^2}y=0
\end{gather}

Now if we can find a suitable function $\theta(z)$ for which $\Bar{\sigma}(z)$ will be divisible by $\sigma(z)$ i.e  $\Bar{\sigma}(z) = \lambda \sigma(z)$ where $\lambda$ is just a constant, then the latter will reduced to the form of a hypergeometric differential equation as follows
\begin{gather} \label{hypergeometrictype}
  \sigma(z) \frac{d^2 y}{dz^2} + \tau(z) \frac{dy}{dz} + \lambda  y(z)=0 
\end{gather}

Putting the value of $\Bar{\sigma}(z)$ into (\ref{sigmabarpi}) we have a quadratic equation in $\pi(z)$ given by
\begin{gather}
    \pi^2(z)- \left( \sigma'(z)-\Tilde{\tau}(z) \right) \pi(z) + \Tilde{\sigma}(z) -k\sigma(z) =0  \quad \text{where } \quad k=\lambda-\pi'(z)
\end{gather} 
where $k$ is a constant as $\pi(z)$ is first degree polynomial as mentioned earlier. Therefore the roots of the above quadratic equation are as follows
\begin{gather} \label{NUpi}
    \pi(z)=\frac{\sigma'(z)-\Tilde{\tau}(z)}{2} \pm \Big[(\frac{\sigma'(z)-\Tilde{\tau}(z)}{2})^2-\Tilde{\sigma}(z)+k\sigma(z)   \Big]^\frac{1}{2}  
\end{gather}

Now to get a plausible and physical solution we need to come up with a particular $\tau(z)$-function which will have a negative derivative. So, for that we need to fix which root to choose for $\pi(z)$ which will consequently determine the suitable $\theta(z)$ function. In that way a common trend is being followed-  find the value of $k$, from (\ref{NUpi}), for which the expression under the square-root becomes a perfect square function of $z$ and then calculate the polynomials $\pi(z)$ and $\tau(z)$.  

After fixing these functions we return to (\ref{hypergeometrictype}) to calculate $y(z)$ and $\lambda$ \cite{nik1988,moh2012,rob2013}. Here we use a property of hypergeometric functions, to evaluate $\lambda$, which tells that all higher order derivative of a hypergeometric function is also a hypergeometric function. Therefore the solution of $(\ref{hypergeometrictype})$ can be generalized and it dictates the $\lambda$ as follows 
\begin{gather} \label{lamda}
    \lambda=k+\pi'(z)=\lambda_n=-n\tau'(z)-\frac{n(n-1}{2} \sigma''(z)
\end{gather}
where $n=0,1,2,.....$ while the solution (which is a hypergeometric function) namely $y(z)$ is given by in terms of Rodrigues formula 
\begin{gather} \label{rodformula}
    y_n(z)= \frac{a_n}{\rho(z)}\frac{d^n}{dz^n}[\sigma^n(z)\rho(z)]
\end{gather}
where $a_n$ is the normalization constant and the weight function $\rho(z)$ suffices the relation $\left(\sigma(z)\rho(z) \right)'=\tau(z)\rho(z)$ which can be rewritten as  
\begin{gather} \label{rho}
    \frac{\rho'(z)}{\rho(z)}=\frac{\tau(z)-\sigma'(z)}{\sigma(z)}
\end{gather}

And finally  we can write the full solution for the equation $(\ref{NUequation})$ as $\Sigma(z)=y_n(z) \theta(z)$.

\section{Data availability statement}

All data supporting the findings of this study are included in the article.


\begin{thebibliography}{}

\bibitem{dirac} P. A. M. Dirac  The principles of quantum mechanics, {\it Clarendron Press, Oxford} (1930)

\bibitem{tha1992} B. Thaller  The Dirac equation, {\it Springer, Singapore} (1992)

\bibitem{kat} M. I. Katsnelson, K. S. Novoselov and A. K. Geim  {\it Nature Phys.} {\bf 2} 620 (2006)

\bibitem{khr} V. V. Khruschov 	{\it arXiv:1004.2116} (2010)

\bibitem{nov} K. S. Novoselov, A. K. Geim, S. V. Morozov, D. Jiang, Y. Zhang, S. V. Dubonos, I. V. Grigorieva and A. A. Firsov  {\it Science} {\bf 306} 666 (2004)

\bibitem{cast} A. H. Castro Neto, F. Guinea, N. M. R. Peres, K. S. Novoselov and A. K. Geim {\it Rev. Mod. Phys.} {\bf 81} 109 (2009)

\bibitem{gall} A. Gallerati {\it Eur. Phys. J. Plus} {\bf 134} 202 (2019)

\bibitem{jakub22} V. Jakubský, Ş. Kuru, and J. Negro
{\it Phys. Rev.} {\bf B 105} 165404 (2022)

\bibitem{fer2020} M. Castillo-Celeita  and D. J. Fernández C. {\it J Phys {\bf A} Math Theor} {\bf 53} 035302 (2020)


\bibitem{lon2010} S. Longhi {\it Phys. Rev. Let.} {\bf 105} 013903 (2010)

\bibitem{flo2021} J. C. Flores {\it Phys. Let.} {\bf A 385} 126987 (2021)

\bibitem{ben1998} C. M. Bender and S. Boettcher {\it Phys. Rev. Lett.} {\bf 80} 5243 (1998)

\bibitem{bag2000} B. Bagchi and R. Roychoudhury {\it J. Phys. {\bf A}: Math. Gen.} {\bf 33} L1 (2000)

\bibitem{ahm2001} Z. Ahmed  {\it Phys. Let.} {\bf A 282} 343 (2001)

\bibitem{moi} N. Moiseyev Nonhermitian Quantum Mechanics, {\it Cambridge University Press} (2011)

\bibitem{baga2016} F. Bagarello, R. Passante and C. Trapani {\it Springer Proceedings in Physics} {\bf 184} (2016)


\bibitem{elg} R. El-Ganainy K. G. Makris, M. Khajavikhan, Z. H. Musslimani, S. Rotter and D. N. Christodoulides {\it Nature Phys.} {\bf 14} 11 (2018)

\bibitem{cruz} K. Zelaya, S. Cruz y Cruz and O. Rosas-Ortiz  {\it Geom. Methods in Phys., Trends in Mathematics; Birkhäuser: Cham, Switzerland} {\bf XXXVIII} 283 (2020)

\bibitem{ard2009} A. Arda and R. Sever {\it Chin. Phys. Let.} {\bf 26} 090305 (2009)

\bibitem{alh2013} A. D. Alhaidari  {\it Phys. Let.} {\bf A 377}  2003 (2013)

\bibitem{bag2020} B. Bagchi and J. Yang  {\it J. Math. Phys.} {\bf 61} 063506 (2020)

\bibitem{gui2008} G. Gui, J. Li and J. Zhong {\it Phys. Rev.} {\bf B 78} 075435 (2008)

\bibitem{dow2016} C. A. Downing  and M. E. Portnoi {\it Phys Rev} {\bf B 94} 165407 (2016)

\bibitem{ley2020} M. Oliva-Leyva , J. E. Barrios-Vargas ,
and G. Gonzalez de la Cruz {\it Phys. Rev.} {\bf B 102}, 035447 (2020)

\bibitem{dow2017} C. A. Downing and M. E. Portnoi  {\it J. Phys.: Condens. Matter} {\bf  29} 315301 (2017)

\bibitem{jia2006} C. S. Jia and A. de S. Dutra {\it J. Phys. {\bf A}: Math. Gen.} {\bf 39} 11877 (2006)

\bibitem{dut2009} V. G. C. S. dos Santos, A. de Souza Dutra and M.B. Hott {\it  Phys. Let.} {\bf A 373}  3401 (2009)

\bibitem{arda10} A. Arda and R. Sever {\it Phys. Scr.} {\bf 82} 065007 (2010)

\bibitem{bagchi05} B. Bagchi, A. Banerjee, C. Quesne and V. M. Tkachuk {\it J. Phys. A: Math. Gen.} {\bf 38} 2929 (2005)

\bibitem{bagchi04}B. Bagchi, P. Gorain, C. Quesne and R. Roychoudhury {\it Mod. Phys. Lett.}  {\bf A 19} 2765 (2004)

\bibitem{znojil12}M. Znojil and G. Lévai {\it Phys. Lett.} {\bf A 376} 3000 (2012)

\bibitem{sever08} R. Sever, C. Tezcan, M. Aktaş and Ö. Yeşiltaş {\it J. Math. Chem.} {\bf 43} 845 (2008)

\bibitem{bagchi2008} B. Bagchi and T. Tanaka {\it Phys. Lett} {\bf A 372} 5390 (2008)

\bibitem{oscar20} O. Rosas-Ortiz {\it Geometric Methods in Physics XXXVIII. Trends in Mathematics. Birkhäuser, Cham.} (2020)

\bibitem{rah2022} R. Ghosh  {\it  J. Phys. {\bf A}: Math. Theor.} {\bf 55} 015307 (2022)

\bibitem{mus2013} O. Mustafa {\it Cent. Eur. J. Phys.} {\bf 11} 4 (2013)

\bibitem{nik1988} A. V. Nikiforov and  V. B. Uvarov  Special Functions of Mathematical Physics,  {\it Birkhauser, Bassel} (1988)

\bibitem{jun2020} G. Junker {\it Eur. Phys. J. Plus.} {\bf  135} 464 (2020)

\bibitem{bag2021} B. Bagchi  and R. Ghosh  {\it J. Math. Phys.} {\bf 62} 072101 (2021)

\bibitem{per2009} N. M. R. Peres  {\it J. Phys. Cond. Matt.}  {\bf 21} 095501 (2009)

\bibitem{has2013} H. Hassanabadi, S. Zarrinkamar and A. A. Rajabi  {\it Phys. Let.} {\bf B 718} 1111 (2013)

\bibitem{kempf95}A. Kempf, G. Mangano, and R. B. Mann
{\it Phys. Rev.} {\bf D 52} 1108 (1995)

\bibitem{yes2014} Ö. Yeşiltaş  {\it J. Math. Phys.}  {\bf 55} 082106 (2014)

\bibitem{moh2012} M. R. Pahlavani  Theoretical Concepts of Quantum Mechanics,  InTech , Croatia (2012)

\bibitem{rob2013} W. Robin   {\it International Mathematical Forum}  {\bf 8} 1455 - 1466 (2013)





\end{thebibliography}
\end{document}